\documentclass[review]{elsarticle}
\usepackage{graphicx}
\usepackage{amsfonts,amsthm,bm}
\usepackage{tabularx}
\usepackage{mathtools}
\usepackage[makeroom]{cancel}
\usepackage{amssymb}
\usepackage{comment}
\usepackage{float}
\usepackage[dvipsnames]{xcolor}
\usepackage{ifpdf}
\usepackage{cite}
\usepackage{hyperref}
\usepackage{enumitem}
\usepackage{algorithm2e}
\usepackage{bbold}
\usepackage{multicol}
\newtheorem{theorem}{Theorem}

\newtheorem{remark}{Remark}

\newtheorem{definition}{Definition}
  
\journal{Applied Mathematical Modeling}

\begin{document}
\begin{frontmatter}
\title{Dynamic Linepack Depletion Models for Natural Gas Pipeline Networks}

\author{Samuel Chevalier\fnref{fn1}}
\ead{schev@mit.edu}

\author{Dan Wu\corref{cor1}\fnref{fn2}}
\ead{danwumit@mit.edu}

\cortext[cor1]{Corresponding author}
\fntext[fn1]{Samuel Chevalier is with the Department of Mechanical Engineering, Massachusetts Institute of Technology.}
\fntext[fn2]{Dan Wu is with the Laboratory for Information and Decision Systems, Massachusetts Institute of Technology.}

\address{Massachusetts Institute of Technology, Cambridge, MA 02139}

\begin{abstract}
Given the critical role played by natural gas in providing electricity, heat, and other essential services, better models are needed to understand the dynamics of natural gas networks during extreme events. This paper aims at establishing appropriate and fast simulation models to capture the slow dynamics of linepack depletion for ideal isothermal natural gas pipeline networks. Instead of solving partial differential equations (PDE) on a large scale, three alternative { implicit} ordinary differential equation (ODE) simulation techniques are derived and discussed. The first one is commonly used in the literature with a slack node assumption. We show that the system of equations associated with this model is degenerate when flux injections are controlled (i.e. specified) at all nodes. { To recover regularity under such a condition, two novel implicit ODE models are proposed, both with different techniques for specifying boundary conditions.} They are { easy to derive and efficient to simulate with standard ODE solvers. More importantly, they} present useful frameworks for analyzing how networks respond to system-wide mass flux imbalances. These techniques offer different alternatives for simulating system dynamics based on how sources and loads are chosen to be modeled, and they are all proven to be { regular} (non-degenerate) in tree-structured networks. These proposed techniques are all tested on the 20-node Belgium network. The simulation results show that the conventional model with the slack node assumption cannot effectively capture linepack depletion under long term system-wide mass flux imbalance, while the proposed models can characterize the network behavior until the linepack is completely depleted.
\end{abstract}


\end{frontmatter}


\section{Introduction}\label{Introduction}

Natural gas continuous to be a growing fuel source among many energy consumption sectors, especially the power generation sector of the electrical grid. In 2018, natural gas (NG) fired power plants produced 35.5\% of the total electricity in the United States~\citep{Tyra:2019}, up from 13\% in 2000~\citep{Johnson:2002}. Since onsite storage of NG fuel is nonexistent for many of these generators~\citep{Ekstrom:2014}, the NG pipeline network (NGPN) serves as a critical transportation and storage infrastructure system in the American energy supply chain. Thus, the economic and reliable operation of NGPNs plays a critical role in the secure and efficient supply of energy to the whole of society. 

It is increasingly evident that the degree of coupling between the electrical power grid and NGPNs is largely increasing \citep{oliver:1999electrification,Tyra:2019}. This strong coupling raises a new challenge for analyzing the propagation of certain types of failures in these systems~\citep{Chertkov:2015}. Accordingly, much work has been done in discussing design procedures for NGPNs \citep{ruan:2009procedure}, co-planning NGPNs and power grids \citep{unsihuay:2010model,zhao:2017coordinated}, optimal operation of NGPNs \citep{misra:2014optimal,manshadi:2018tight}, impacts of failures and congestion on NGPNs \citep{tran:2018impact,liu:2014look}, and the interdependency of NGPNs and power grids \citep{gil:2015electricity,he:2018robust}. To better understand and mitigate failure cascades between NGPNs and the power grid, dynamic and steady state models utilizing different assumptions and time scales are required. A vast literature exists on the modeling of electrical power systems and their various dynamical components~\citep{Kundur:1994,Sauer:2006,Glover:2007}. When considering long term phenomena, the fast electromagnetic wave transients in power grids can be simplified to algebraic constraints, but the dynamics of NG propagation are inherently slow ($10$ m/s is a characteristic gas flow speed in the pipeline, usually below $20$ m/s) and thus cannot be ignored. 

A variety of platforms have been developed for the purpose of simulating NGPN dynamics. Early attempts and reviews of the relevant NGPN dynamics, modeling strategies, and numerical solution techniques can be found in~\citep{van:1983modelling,Osiadacz:1984,Thorley:1987,osiadacz:1989comparison}. Recent works have focused on either deriving more accurate simulation models with non-isothermal gas assumption \citep{abbaspour:2010transient,lopez:2016steady,chaczykowski:2010transient} or approximating network dynamics through discretization in time and space~\citep{herran:2009modeling,Mak:2019,Uilhoorn:2014,Wen:2018}, so that state space control and optimization techniques can be leveraged for various objectives. For high fidelity results, authors in~\citep{Gyrya:2019} employ a second order staggered finite difference discretization method. The approach is computationally efficient, unconditionally stable, and proven to exactly satisfy mass conservation. Despite the potential high fidelity of these results, simulating the nonlinear partial differential equations (PDEs) which model these fluid dynamics can be intractably slow for large NGPNs whose dynamics unfold over the course of many hours. It is therefore necessary to have fast and relatively accurate tools for evaluating the dynamic behaviour of NGPNs under extreme situations, such as severe contingencies.
	
Simulation speed has been the primary focus of other researchers. By linearizing about an operating point,~\citep{Zhou:2017} employs transfer function matrices to describe the dynamics associated with pipelines in a network. For further simplification,~\citep{Nejad:2008} identifies the dominant eigenmodes of the linearized NGPN and uses them to build a reduced order model of the system for simplified analysis. In order to simplify the particular gas wave effects while still preserving nonlinearities,~\citep{Anatoly:2016} analytically integrates across segments of the pipelines in order to build an ODE model. Via adaptive time-stepping, simulation results of the reduced model are compared with those from a full order PDE model and are shown to be comparable in accuracy. 
For increasing the speed of simulation,~\citep{Grundel:2014} models a NGPN with a set of DAEs, and then model order reduction, via proper orthogonal decomposition, is {proposed for fast evaluations.} {By applying the projector based analysis of linear DAEs~\citep{lamour:2011}, the DAE model in \citep{Grundel:2014} is shown to be at most index-2. Although the projector method ~\citep{lamour:2011,lamour:2011treatment} and the derivative array approach \citep{campbell:1987} can be used to find the solution of high-index DAEs, they are mainly developed for linear DAEs. Their nonlinear application inevitably involves Jacobian function evaluation and kernel manipulation at each time step, which are computationally expensive.} In order to exploit the analytical structure of the equations,~\citep{Qiu:2018} applies a finite volume method to convert the PDEs into relevant ODEs. After eliminating the maximum number of algebraic constraints, a preconditioner is developed in order to reduce the computation burden of Jacobian inversion during each Newton iteration. The proposed methods are shown to significantly speed up backward Euler integration of the network dynamics~\citep{Qiu:2018}. 

While there are many sophisticated methods available for simulating intra-pipeline dynamics, the effects of boundary constraints (sources and loads) have been seldom addressed in the literature. Most (if not all) simulation platforms {with reduced ODE models} assume constant pressure at a slack node, but this further assumes the network has an infinite pool of reserves to pull from at infinitely fast injection rates. While this may be a safe assumption during ``normal" system operation, it could certainly fail during certain critical contingencies. In this paper, we investigate alternative methods for applying different boundary constraints {such that no infinite pool of reserves is needed at slack nodes. These proposed methods can be very useful when long term imbalances of mass flow rates exist in the pipeline network. Such imbalance will result in the so called ``linepack depletion'' phenomenon.} Our primary goal is to develop {implicit regular ODE models which are easy to implement and can} better capture the dynamic characteristics of the linepack depletion phenomenon for extreme contingency situations. Specifically, we require our models to provide an accurate estimation of the survival time during linepack depletion. Thus, rather than the exact transient behavior of the NG, we are primarily concerned with developing {computationally efficient} methods which can {appropriately} characterize the linepack depletion in the network over time. When these methods are posed properly, simulation results can be collected on a timescale which can enable analyses of cascading failures in coupled NG and electrical power systems. With a focus on developing fast and well-characterized simulation techniques, the specific contributions of this paper are as follows:

\begin{enumerate}

\item We leverage the work presented in~\citep{Anatoly:2016,Grundel:2014} by building a full network model on top of the previously proposed framework. This model is proved to be degenerate {(the canonical subspace of the implicit ODE does not expand to the entire Euclidean state space)} if mass flux injections are specified at all boundary points. 

\item After presenting a common solution to this degeneracy (introduction of a constant pressure slack node), we derive two alternative { implicit regular} ODE simulation models each of which make different load and source modeling assumptions. We prove that they are { regular (non-degenerate)} and explain the particular applications of each.

\item We compare and contrast simulation results associated with these different simulation models for a given set of contingencies. Most notably, we highlight the inability of the common slack node model to capture linepack depletion effects when NG source injections are constrained.

\end{enumerate}

 The remainder of the paper is structured as follows. In Section \ref{Network Model}, we construct a NGPN model and prove model degeneracy for certain load and source modeling conditions. In Section \ref{Simulation Techniques}, we overcome this degeneracy by presenting three alternative simulation techniques associated with various load and source modeling assumptions. In Section \ref{Test Results}, we present and compare test results associated with the derived models.


\section{Building an ODE Simulation Platform}\label{Network Model}
In this section, we build the relevant set of ODEs which are used for simulating gas network transients.

\subsection{The Euler Equations for Ideal Isothermal Gas}
We begin by stating the Euler equations~\citep{Anatoly:2016} which govern unsteady compressible ideal isothermal fluid flow in one dimension ($x$), where all relevant variables and constants are explained in Table \ref{tab:vars}:
\begin{align}
\frac{\partial\rho}{\partial t}+\frac{\partial}{\partial x}(\rho\nu)&=0\label{eq: mass}\\
\frac{\partial}{\partial t}\left(\rho\nu\right)+\frac{\partial}{\partial x}\left(\nu^{2}\rho\right)+\frac{\partial}{\partial x}p & =-\frac{\lambda}{2D}\rho\nu|\nu|-\rho g\sin(\theta).\label{eq: momentum}
\end{align}

\begin{table}[ht!]
\renewcommand{\arraystretch}{1}
\caption{Gas Flow Variables and Constants}
\centering\label{tab:vars}
\begin{tabular}{|c|c|c|}
\hline
\textbf{\;\;Symbol\;\;} & \textbf{\;\;\;\;\;\;\;\;\;\;\;\;\;\;\;\;Variable\;\;\;\;\;\;\;\;\;\;\;\;\;\;\;\;} & \textbf{\;\;\;\;Units\;\;\;\;}\\
\hline
\hline
$x$       & Distance               & ${\rm m}$\\
$t$       & Time                   & ${\rm s}$\\
$\phi$    & Mass Flux Flow         & ${\rm kg}/{{\rm m}^2\!\cdot\!{\rm s}}$\\
$d$       & Mass Flow         & ${\rm kg}/{\rm s}$\\
$\rho$    & Gas Density            & ${\rm kg}/{{\rm m}^3}$\\
$\nu$     & Gas Velocity           & ${\rm m}/{\rm s}$\\
$p$       & Gas Pressure           & ${\rm N}/{{\rm m}^2}$\\
$T$       & Temperature            & ${\rm K}$\\
$R$       & Ideal Gas Constant     & ${\rm J}/{\rm K\!\cdot\!mol}$\\
$D$, $L$  & Pipe Diameter, Length  & ${\rm m}$\\
$A$       & Pipe Area              & ${\rm m}^2$\\
$g$       & Gravity                & ${\rm m}/{{\rm s}^2}$\\
$\theta$  & Pipe Angle             & ${\rm rad}$\\
$\alpha$  & Compressor Ratio       & $-$\\
$Z$       & Compressibility Factor & $-$\\
$\lambda$ & Darcy Friction Factor  & $-$\\
\hline
\end{tabular}
\end{table}Equation (\ref{eq: mass}) is a statement of the continuity of mass flow while (\ref{eq: momentum}) is a statement of the conservation of momentum. They are commonly used to describe the dynamics of an ideal gas. If we further assume an isothermal gas (i.e. constant temperature), then there is no energy conservation equation and there exists a linear relationship between pressure $p$ and density $\rho$:
\begin{equation}
    p=a^2\rho,\label{eq: a^2}
\end{equation}
where $a^2=ZRT/M$ comes from the ideal gas law (plus some constant nonideality correction factor $Z$). Further assuming the gravitation forces caused by elevation changes are negligible ($\theta\approx 0$), and that fluid flows are much slower than the speed of sound ($\nu\ll a$), the PDEs simplify to
\begin{align}\label{eq: s1}
\frac{\partial}{\partial t}\rho+\frac{\partial}{\partial x}\phi&=0\\
\frac{\partial}{\partial t}\phi+a^{2}\frac{\partial}{\partial x}\rho & =-\frac{\lambda}{2D}\phi\frac{|\phi|}{\rho}.\label{eq: s2}
\end{align}

\noindent Eqn \eqref{eq: s1}-\eqref{eq: s2} are the first order linear hyperbolic PDEs which can take the following form:

\begin{equation}
    \frac{\partial}{\partial t} 
    \begin{bmatrix}
         \rho\\[\jot]
        \phi
    \end{bmatrix}+
    A \frac{\partial}{\partial x}
    \begin{bmatrix}
         \rho\\[\jot]
        \phi
    \end{bmatrix} = f(\rho,\phi)\label{eq:1st_PDE}
\end{equation}%
where $A=\begin{bmatrix}
        0~~1\vspace{-3pt}\\
        a^2~0
    \end{bmatrix}$ and $f(\rho,\phi)=\begin{bmatrix}
        0\vspace{-3pt}\\
        -\frac{\lambda}{2D}\phi\frac{|\phi|}{\rho}
    \end{bmatrix}$.
Since the eigenvalues of matrix $A$ in Eqn~\eqref{eq:1st_PDE} are $\pm a$, the characteristic families propagate in the opposite directions on a finite interval, say, $x \in [0,L]$. Thus, the well-posedness of Eqn~\eqref{eq:1st_PDE} requires boundary conditions on different ends of the interval for state variables $\rho$ and $\phi$, respectively. These boundary conditions can be either $\big(\rho(t,0),\phi(t,L)\big)$ or $\big(\rho(t,L),\phi(t,0)\big)$ \citep{strikwerda:2004finite}.

Failure to provide these boundary conditions can yield potentially ill-posed natural gas PDE models, which is discussed in~\citep{Chaczykowski:2017}. While existing literature has extensively discussed the well-posedness of general hyperbolic PDEs \citep{secchi:1995linear,secchi:1996well,coulombel:2005well}, the boundary conditions of its reduced ODE lumped counterpart has seldom been addressed for the natural gas network. One theme of this paper is to reveal the ill-posedness of the reduced ODE lumped model under certain boundary conditions which may be encountered in severe contingencies, and come up with novel techniques to solve these issues.

\subsection{Model Reduction of the Euler Equations}
We now consider some line of length $L$. This line is spatially discretized into $N$ segments, each of length $l=L/N$. Intermediate nodes are thus defined at  $x=\{0,\,l,\,2l... Nl=L\}$. As discussed in~\citep{Anatoly:2016}, the mass and momentum equations can be approximately integrated over the length of the line segment $x\in[0,\,l]$ through the trapezoidal rule, reducing the the PDEs to a set of equations only related to time derivatives. 
The resulting nonlinear differential equations contain derivatives with respect to time only:
\begin{align}
\frac{l}{2}\left(\dot{\rho}_{l}\!+\!\dot{\rho}_{0}\right)+\left(\phi_{l}\!-\!\phi_{0}\right)&=0\label{eq: mass_int}\\
\frac{l}{2}(\dot{\phi}_{l}\!+\!\dot{\phi}_{0})+a^{2}\left(\rho_{l}\!-\!\rho_{0}\right)&=\frac{-l\lambda}{4D}\left(\phi_{l}\!+\!\phi_{0}\right)\frac{|\phi_{l}\!+\!\phi_{0}|}{\rho_{l}\!+\!\rho_{0}},\label{eq: momentum_int}
\end{align}
where subscripts $0$ and $l$ indicate variables corresponding to the beginnings and ends of line segments, respectively. Via this integration, \citep{Anatoly:2016} confirms the validity of this reduction technique for a slowly varying input on a small test system.
\begin{remark}
    The accuracy associated with the model reduction in (\ref{eq: mass_int})-(\ref{eq: momentum_int}) can be increased arbitrarily if we integrate on shorter cascaded segments of the line, thus driving $l \rightarrow 0$.
\end{remark}

There are various other integration/simplification schemes to convert natural gas PDE models into ODE/DAE ones. Interested readers can refer to \citep{Wen:2018,Gyrya:2019,Zhou:2017,Nejad:2008,Qiu:2018}. In discussions of this paper, we only focus on the ill-posed nature of this specific reduced model, leaving other interesting models for the future investigation.

To specify a complete network model with an arbitrary number of lines connecting sources and loads, these equations must be complemented by two others. The first is a statement of the conservation of mass flux at each node; the second states that the intra-nodal pressures on the incoming and outgoing sides of a node are algebraically related according to some compressor amplification factor $\alpha$. If no compressor is present at a node, $\alpha=1$. This parameter may be independently controlled by the system operators as necessary and is treated as a system input. 

\subsection{NGPN Simulation Model}
We now set up a system of equations to describe the dynamics of a full NGPN. Our primary goal is to investigate the phenomenon of linepack depletion. Accordingly, in writing the conservation laws, we assume that the load and source mass flux injections are specified inputs (typically constant). We also assume all density states are free variables, i.e., they may evolve freely according to the dynamics of the system. 
\begin{definition}
    The NG simulation model where all mass flux injections are specified (i.e. determined) and all density states are free variables is referred to as the \textbf{flux determined} model.
\end{definition}
\noindent As shall be shown, the differential system associated with the flux determined system is degenerate, {which requires complicated transformation and projection techniques to reveal the dynamics on the canonical subspace (canonical submanifold if nonlinear)~\citep{lamour:2011}. In order to reduce the simulation complexity,} alterations should be made.

We define an arbitrary network with $\hat n$ physical nodes and $\hat m$ physical branches. Once the system lines have been properly discretized, we include the intermediate nodes and branches to define a total of $n\gg{\hat n}$ nodes and $m\gg{\hat m}$ branches. The standard signed incidence matrix related to the system is given by $E\in\mathbb{R}^{m\times n}$. The graph is directed in the same direction in which line flows are normally positive. We define two vectors: $\boldsymbol{\overline{\rho}}\in\mathbb{R}^{2n\times1}$ and $\overline{\boldsymbol{\phi}}\in\mathbb{R}^{2m\times1}$. The vector $\overline{\boldsymbol{\rho}}$ is filled with the density states on either ``side" of each of the $n$ nodes, where ``$-$" indicates the inflowing side and ``$+"$" indicates the outflowing side of each node\footnote{Any incoming edge to node $i$ will have density $\rho_-^{(i)}$ at the point of interconnection, and outgoing edge will have density $\rho_+^{(i)}$ at the point of interconnection.}. The vector $\overline{\boldsymbol{\phi}}$ is filled with the mass flux states on either side of each of the $m$ lines, where ``$0$" indicates the beginning side and ``$l$" indicates the ending side of each line:
\begin{equation*}
\overline{\boldsymbol{\rho}} =\left[\begin{array}{c}
\rho_{-}^{(1)}\\
\vdots\\
\rho_{-}^{(n)}\\
\rho_{+}^{(1)}\\
\vdots\\
\rho_{+}^{(n)}
\end{array}\right],\;\;\;
\overline{\boldsymbol{\phi}} =\left[\begin{array}{c}
\phi_{0}^{(1)}\\
\vdots\\
\phi_{0}^{(m)}\\
\phi_{l}^{(1)}\\
\vdots\\
\phi_{l}^{(m)}
\end{array}\right].
\end{equation*}

\noindent Next, we split these vectors in half and define subset vectors:
\begin{equation*}
\boldsymbol{\rho}_{-}\!=\!\left[\!\!\begin{array}{c}
\rho_{-}^{(1)}\\
\rho_{-}^{(2)}\\
\vdots\\
\rho_{-}^{(n)}
\end{array}\!\!\right]\!\!,\;\;\boldsymbol{\rho}_{+}\!=\!\left[\!\!\begin{array}{c}
\rho_{+}^{(1)}\\
\rho_{+}^{(2)}\\
\vdots\\
\rho_{+}^{(n)}
\end{array}\!\!\right]\!\!,\;\;\boldsymbol{\phi}_{0}\!=\!\left[\!\!\!\begin{array}{c}
\phi_{0}^{(1)}\\
\phi_{0}^{(2)}\\
\vdots\\
\phi_{0}^{(m)}
\end{array}\!\!\!\right]\!\!,\;\;\boldsymbol{\phi}_{l}\!=\!\left[\!\!\!\begin{array}{c}
\phi_{l}^{(1)}\\
\phi_{l}^{(2)}\\
\vdots\\
\phi_{l}^{(m)}
\end{array}\!\!\!\right]\!\!.
\end{equation*}
We define $\boldsymbol{\alpha}\in\mathbb{R}^{n\times1}$ as the vector of compressor ratios. The compressors relate the density\footnote{Since compressor ratios linearly relate nodal pressures, nodal densities are also related by the same ratios.} differentials via
\begin{align}\label{eq: comp_ratios}
\boldsymbol{\rho}_{+} & =\text{diag}\left\{ \boldsymbol{\alpha}\right\} \boldsymbol{\rho}_{-}.
\end{align}
\begin{figure}
\centering
\includegraphics[width=0.6\textwidth]{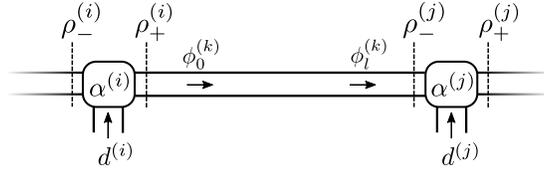}
\caption{Shown are the density, mass flow injection, and compressor variables and constants associated with nodes $i$ and $j$. Also shown are the mass flux flow variables and their respective reference directions on line $k$, where line $k$ connects nodes $i$ and $j$. If these are physical nodes, then $l=L$. Otherwise, $l$ represents the length of the line discretization proposed in (\ref{eq: mass_int})-(\ref{eq: momentum_int}), and the nodes are induced ``intermediate" nodes. \label{eq: Pipeline_Example}}
\end{figure}Fig. \ref{eq: Pipeline_Example} illustrates the relationships between compressor constants and flux injection, flux line flow, and density variables. 
For notation convenience, we introduce the density variable vector $\boldsymbol{\rho}\equiv\boldsymbol{\rho}_{-}$. 

Next, we must define a mass-flow conservation law. In this network, just as in power systems, a positive injection is defined to be a source of supply, while a negative injection represents a load. We define the load and source injection vector $\boldsymbol{d}\in\mathbb{R}^{n\times1}$ whose relationship satisfies
\begin{align}\label{eq: conservation_network}
K_{0}\boldsymbol{\phi}_{0}+K_{l}\boldsymbol{\phi}_{l} & =\boldsymbol{d}.
\end{align}
Matrices $K_{l}\in\mathbb{R}^{n\times m}$ and $K_{0}\in\mathbb{R}^{n\times m}$ codify, respectively, which lines enter and leave each node (based on the convention of the incidence matrix):
\begin{align}\label{eq: K0}
K_{0}(i,j) & =\begin{cases}
A_j, & \;\;\,\text{Line} \; j \; \text{leaves node} \; i\\
0, & \;\;\,\text{Line} \; j \; \text{does not leave node} \; i
\end{cases}\\
K_{l}(i,j) & =\begin{cases}
-A_j, & \text{Line} \; j \; \text{enters node} \; i\\
0, & \text{Line} \; j \; \text{does not enter node} \; i,
\end{cases}\label{eq: KL}
\end{align}
where $A_j$ is the cross-sectional area of line $j$. Equation (\ref{eq: conservation_network}) is a linear algebraic equation which is entirely analogous to Kirchhoff's Current Law in an electrical network. When taking the time derivatives of Eqn \eqref{eq: comp_ratios} and \eqref{eq: conservation_network}, we have
\begin{align}\label{eq: comp_conserv}
\dot{\boldsymbol{\rho}}_{+} =\text{diag}\left\{ \boldsymbol{\alpha}\right\} \dot{\boldsymbol{\rho}}_{-}\\
K_{0}\dot{\boldsymbol{\phi}}_{0}+K_{l}\dot{\boldsymbol{\phi}}_{l} =\dot{\boldsymbol{d}}.
\end{align}

\begin{remark}
    While (\ref{eq: conservation_network}) represents the conservation of mass flow at each node in the network, (\ref{eq: mass}) represents the continuity of differential mass flow on a pipeline. These are physically similar, yet characteristically different, processes.
\end{remark}
For notation convenience, we also define $\overline{K}_{0}$ and $\overline{K}_{L}$, which are equivalent to $(\ref{eq: K0})-(\ref{eq: KL})$, but un-scaled by the cross-section areas. Stated succinctly,
\begin{align}
\overline{K}_{0} & = \tfrac{1}{2}\big(|E|^{\top}+E^{\top}\big)\\
\overline{K}_{l} & = \tfrac{1}{2}\big(E^{\top}-|E|^{\top}\big),
\end{align}%
where, again, $E$ is the signed incidence matrix.

Using the proposed notation, (\ref{eq: mass_int}) may be written as 
\begin{subequations}
\begin{align}
{\bf 0} & ={\rm diag}\left\{{\bf l}/2\right\} \left(\overline{K}_{0}^{\top}\dot{\boldsymbol{\rho}}_{+}-\overline{K}_{l}^{\top}\dot{\boldsymbol{\rho}}_{-}\right)+\left(\boldsymbol{\phi}_{l}-\boldsymbol{\phi}_{0}\right)\\
 & =\underbrace{{\rm diag}\left\{{\bf l}/2\right\} \left(\overline{K}_{0}^{\top}\text{diag}\left\{ \boldsymbol{\alpha}\right\} -\overline{K}_{l}^{\top}\right)}_{\Gamma_{1}}\dot{\boldsymbol{\rho}}+\boldsymbol{\phi}_{l}-\boldsymbol{\phi}_{0}.
\end{align}
\end{subequations}
where ${\bf l} \in {\mathbb R}^{m\times 1}$ is a vector of line lengths. In considering (\ref{eq: momentum_int}), we note the Hadamard product $\odot$ which performs element by element multiplication on a set of vectors, and Hadamard division $\oslash$ which performs element by element division on a set of vectors. We also define function $\bf f$, which is a function of identically sized vectors $\bf a$, $\bf b$, and $\bf c$:
\begin{align}
    {\bf f}({\bf x},{\bf y},{\bf z}) \coloneqq \left({\bf x}+{\bf y}\right)\odot\left|{\bf x}+{\bf y}\right|\oslash{\bf z}.
\end{align}
We also define a set of $\Gamma$ matrices:
\begin{align}
\Gamma_{2} & ={\rm diag}\left\{ {\bf l}/2\right\} \\
\Gamma_{3} & ={\rm diag}\left\{ {\bf a}\right\} ^{2}\left({\overline K}_{l}^{\top}+{\overline K}_{0}^{\top}\text{diag}\left\{ \boldsymbol{\alpha}\right\}\right) \\
\Gamma_{4} & ={\rm diag}\left\{ {\bf l}\odot\boldsymbol{\lambda}\oslash{\bf D}/4\right\}\\
\Gamma_{5} & ={\overline K}_{0}^{\top}\text{diag}\left\{ \boldsymbol{\alpha}\right\} -{\overline K}_{l}^{\top},
\end{align}
where ${\bf a}\in{\mathbb R}^{m\times1}$ is the vector of velocity parameters from (\ref{eq: a^2}), ${\boldsymbol \lambda}\in{\mathbb R}^{m\times1}$ is the vector of darcy friction factors, and ${\bf D}\in{\mathbb R}^{m\times1}$ is the vector of pipe diameters. Thus, (\ref{eq: momentum_int}) can be written as
\begin{subequations}\label{eq: network_momentum}
\begin{align}
\Gamma_2(\dot{\boldsymbol{\phi}}_{l}+\dot{\boldsymbol{\phi}}_{0}) &={\rm diag}\left\{ {\bf a}\right\} ^{2}\left({\overline K}_{l}^{\top}\boldsymbol{\rho}_{-}+{\overline K}_{0}^{\top}\boldsymbol{\rho}_{+}\right)\\
  -\Gamma_4 (\boldsymbol{\phi}_{l}&+\boldsymbol{\phi}_{0})\odot|\boldsymbol{\phi}_{l}+\boldsymbol{\phi}_{0}|\oslash\left({\overline K}_{0}^{\top}\boldsymbol{\rho}_{+}-{\overline K}_{l}^{\top}\boldsymbol{\rho}_{-}\right)\nonumber\\
 & = \Gamma_{3}\boldsymbol{\rho}-\Gamma_{4}{\bf f}\left(\boldsymbol{\phi}_{l},\boldsymbol{\phi}_{0},\Gamma_{5}\boldsymbol{\rho}\right).
\end{align}
\end{subequations}
We may now assemble the full set of differential equations which are necessary to describe the dynamics of this network:
\begin{align}
K_{l}\dot{\boldsymbol{\phi}}_{l}+K_{0}\dot{\boldsymbol{\phi}}_{0} & =\dot{\boldsymbol{d}}\label{eq: de1}\\
\Gamma_{1}\dot{\boldsymbol{\rho}} & =\boldsymbol{\phi}_{0}-\boldsymbol{\phi}_{l}\label{eq: de2}\\
\Gamma_{2}(\dot{\boldsymbol{\phi}}_{l}+\dot{\boldsymbol{\phi}}_{0}) & =\Gamma_{3}\boldsymbol{\rho}-\Gamma_{4}{\bf f}\left(\boldsymbol{\phi}_{l},\boldsymbol{\phi}_{0},\Gamma_{5}\boldsymbol{\rho}\right).\label{eq: de3}
\end{align}
Next, we define the state variable vector ${\bf x}\in\mathbb{R}^{(n+2m)\times1}$:
\begin{align}\label{eq: s_vars}
{\bf x} & =\left[\begin{array}{c}
\boldsymbol{\rho}\\
\boldsymbol{\phi}_{0}\\
\boldsymbol{\phi}_{l}
\end{array}\right],
\end{align}
and we define the RHS of the set of differential equations (\ref{eq: de1})-(\ref{eq: de3}) as ${\bf G}({\bf x})$, such that
\begin{align}
\underbrace{\left[\begin{array}{ccc}
{\bf 0} & K_{0} & K_{L}\\
\Gamma_{1} & {\bf 0} & {\bf 0}\\
{\bf 0} & \Gamma_{2} & \Gamma_{2}
\end{array}\right]}_{M_{{\bf x}}}\dot{{\bf x}}={\bf G}({\bf x}).\label{eq: compact_ode}
\end{align}
The full differential model is compactly stated $M_{{\bf x}}\dot{{\bf x}}={\bf G}({\bf x})$.

\begin{remark}
    The differential order of (\ref{eq: compact_ode}) may be reduced by eliminating the out-flowing flux variable at all nodes (vitual or real) with a single in-flowing flux variable (or vice versa). In this case, $\phi_l^k\equiv\phi_0^{k+1}$. This is similarly noted in~\citep{Qiu:2018}.
\end{remark}


Before concluding this subsection, we define a NG pipeline system $\Sigma_c$ which has a set of specific properties.
\begin{definition}
    Consider a NG pipeline system $\Sigma_c$ whose graph ${\mathcal G}({\mathcal V},{\mathcal E})$ has edge (line) set $\mathcal{E}$, $|\mathcal{E}|=m$, vertex (node) set $\mathcal{V}$, $|\mathcal{V}|=n$, and directed nodal incidence matrix $E\in{\mathbb R}^{m\times n}$. The graph of $\Sigma_c$ has a connected tree structure, meaning $n = m+1$, and the dynamics of the network are codified by (\ref{eq: compact_ode}).
\end{definition}

\subsection{Degeneracy of the Flux Determined NG Simualtion Model}
Now that a differential model for the flux determined model has been derived, we may investigate its shortcomings.
\begin{theorem}\label{theorem: singularity}
    Consider system $\Sigma_c$. The associated coefficient mass matrix $M_{\bf x}$ is a singular matrix.
    \begin{proof}
    Since the columns of $\Gamma_1$ are surrounded by $\bf 0$ on the top and bottom in (\ref{eq: compact_ode}), degenerate column rank of $\Gamma_1$ implies degenerate column rank of $M_{{\bf x}}$, further implying matrix singularity. Since $\Gamma_1\in{\mathbb R}^{m\times n}$, and $m<n$, the column rank of $\Gamma_1$ must be degenerate.
    \end{proof}
\end{theorem}
Theorem~\ref{theorem: singularity} implies that
{ the implicit ODE model \eqref{eq: compact_ode} is degenerate. So the solution of \eqref{eq: compact_ode} should lie in the canonical subspace (canonical submanifold in nonlinear case) which is only a submanifold of the Euclidean state space \citep{lamour:2011}. This suggests that an arbitrary initial point is not necessarily consistent with the canonical subspace. Thus, the resulting trajectory may not be well-defined.} 

In the case where a load perturbation occurs, it would be physically meaningful to eliminate the load density and treat it as an algebraic variable which can instantaneously respond to the injection change; this observation will inform one of our proposed simulation formulations, along with the following theorem.
\begin{definition}
    Consider the signed incidence matrix $E\in{\mathbb R}^{m\times (m+1)}$ of a tree network. The \textit{reduced} incidence matrix $E'\in{\mathbb R}^{m\times m}$ is equal to $E$, but with one column (any column deleted). Square matrix $E'$ is known to be full rank.
\end{definition}

\begin{theorem}\label{theorem: delete_pressure}
    Consider system $\Sigma_c$. To achieve full column rank of $M_{\bf x}$, it is sufficient to delete a density state.
    \begin{proof}
    For full column rank of $M_{\bf x}$ to be achieved, full column rank of $\Gamma_1$ is necessary. To achieve full column rank of $\Gamma_1$, one of its columns must be eliminated, per the proof of Theorem \ref{theorem: singularity}. All columns of $\Gamma_1$ in (\ref{eq: compact_ode}) are associated with a density state variable. The deletion of a column from $\Gamma_1$ will render a square matrix termed $\Gamma_1'$. Matrix $\Gamma_1'$ is structurally equivalent to a scaled version of the system's reduced incidence matrix. Since the reduced incidence matrix of a tree network is full rank, then so is $\Gamma_1'$.
    \end{proof}
\end{theorem}



\section{Simulation Techniques}\label{Simulation Techniques}
With the observations of Theorems \ref{theorem: singularity} and \ref{theorem: delete_pressure} in mind, this section introduces three alternative {implicit regular ODE simulation models} based on different assumptions about the nature of the sources and loads in the system.

\subsection{Technique 1: Infinite Flux Reservoir}
This modeling technique holds at least one pressure state in the system constant, and it has been implicitly employed in other works~\citep{Mak:2019,Anatoly:2016,lamour:2011}. Its framework may be a poor one for investigating linepack depletion, though. Because a density state variable must be eliminated, we must delete a corresponding equation in order to ensure that the number of equations continues to match the number of state variables. The following theorem shows how non-singularity of the mass matrix can be achieved.

\begin{theorem}\label{theorem: delete_conservation}
    Consider system $\Sigma_c$. The elimination of a pressure state and a flux conservation equation in (\ref{eq: de1}) from $M_{\bf x}$ will yield a square, full rank matrix $M_{\bf x}''$.
    \begin{proof}
        Once a pressure state has been eliminated, submatrix $\Gamma_1'$ will be full rank. We define 
        \begin{align}
           M'_{{\bf x}} = \left[\begin{array}{ccc}
            {\bf 0} & K_{0} & K_{L}\\
            \Gamma_{1}' & {\bf 0} & {\bf 0}\\
            {\bf 0} & \Gamma_{2} & \Gamma_{2}
            \end{array}\right].\label{eq: Mxp}
        \end{align}
       We eliminate one row from the top subsection of $M'_{{\bf x}}$ (from matrices ${K}_{0}$ and $K_{L}$) and consider the rank of the submatrix ${\overline M}_{{\bf x}}''$, where
        \begin{align}
           {\overline M}_{{\bf x}}'' = \left[\begin{array}{cc}
            { K}_{0}'      & { K}_{L}'\\
            \Gamma_{2} & \Gamma_{2}
            \end{array}\right].\label{eq: M_bpp}
        \end{align}
        and the prime notation on ${ K}_{0}'$ and ${ K}_{L}'$ indicates the deletion of a row. We take the determinant of ${\overline M}_{{\bf x}}''$:
        \begin{subequations}
        \begin{align}
            {\rm det}\left(\overline{M}_{{\rm {\bf x}}}''\right) & ={\rm det}\left({K}_{0}'-{K}_{L}'\Gamma_{2}^{-1}\Gamma_{2}\right){\rm det}\left(\Gamma_{2}\right)\\
             & ={\rm det}\left({K}_{0}'-{K}_{L}'\right)\sigma.
        \end{align}
        \end{subequations}
        where $\sigma\ne 0$ since $\Gamma_{2}$ is a diagonal matrix. Since $K_{0}'-K_{L}'$ is equal to the scaled, unsigned incidence matrix of a tree network with with one node deleted, it is square and necessarily full rank. Therefore, ${\rm det}(\overline{M}_{{\rm {\bf x}}}'')\propto{\rm det}(K_{0}'-K_{L}')\ne0$. Because ${\overline M}_{{\bf x}}''$ and $\Gamma_1'$ are both full rank, it implies that matrix ${M}_{{\bf x}}''$, which is defined by 
        \begin{align}\label{eq: Mxpp}
           M''_{{\bf x}} = \left[\begin{array}{ccc}
            {\bf 0} & K_{0}' & K_{L}'\\
            \Gamma_{1}' & {\bf 0} & {\bf 0}\\
            {\bf 0} & \Gamma_{2} & \Gamma_{2}
            \end{array}\right],
        \end{align}
        is also full rank.
    \end{proof}
\end{theorem}

Therefore, the elimination of a conservation equation from (\ref{eq: de1}) and the elimination of a density state from (\ref{eq: s_vars}) will yield a set of equations which can be uniquely simulated. When this happens, the density state becomes a model input, and the injection variable $d$ associated with the eliminated conservation equation is also eliminated. When the density state and conservation equation are eliminated at the same node, this node then has the interpretation of a so-called ``slack" node, whose definition follows.
\begin{definition}
A \textbf{slack node} is a node whose density (i.e. pressure) is specified and whose corresponding flux injection can take any unbounded instantaneous value to meet the specified density constraint. 
\end{definition}
The assumption behind the slack node is that it has the ability to pull from an infinite flux reservoir. This slack node is entirely analogous to the infinite bus in power system simulations, where complex power injection can take any value such that a specified complex voltage value is met.

{A similar result to Theorem~\ref{theorem: delete_conservation} has been suggested in \citep{Grundel:2014}, claiming that the DAE system is index-1 if only one supply (slack) node exists in the network. This statement is equivalent to the regularity condition of our implicit ODE model \eqref{eq: compact_ode}. However, in \citep{Grundel:2014}, compressors are not modeled; while in our ODE model \eqref{eq: compact_ode} general compressors are explicitly considered.}

\subsubsection*{Simulation Model Statement}
To implement this simulation technique, we assume without loss of generality (WLOG) that the first node in $\Sigma_c$ will be treated as a slack node. We split system $\bf G$ into the to-be eliminated conservation equation ${\bf G}_1$ and the remaining equations $\tilde {\bf G}$:
\begin{equation}
{\bf G}=\left[\begin{array}{c}
{\bf G}_{1}\\
\hline \tilde{{\bf G}}
\end{array}\right].
\end{equation}
We also split state variable vector $\bf x$ into the to-be eliminated density variable $\rho^{(1)}$ and the remaining variables $\tilde {\bf x}$:
\begin{equation}\label{eq: xtilde}
{\bf x}=\left[\begin{array}{c}
\rho^{(1)}\\
\hline \tilde{{\bf x}}
\end{array}\right].
\end{equation}
We parse the system according to
\begin{equation}\label{eq: parsed_sys}
\left[\begin{array}{c|c}
M_{x1} & M_{x2}\\
\hline M_{x3} & M_{x4}
\end{array}\right]\left[\begin{array}{c}
\dot{\rho}^{(1)}\\
\hline \dot{\tilde{{\bf x}}}
\end{array}\right]=\left[\begin{array}{c}
{\bf G}_{1}(\rho^{(1)},\tilde{{\bf x}})\\
\hline \tilde{{\bf G}}(\rho^{(1)},\tilde{{\bf x}})
\end{array}\right],
\end{equation}
and we eliminate the top equation in (\ref{eq: parsed_sys}). Finally, we parse the second equation, such that
\begin{equation}
M_{x4}\dot{\tilde{{\bf x}}}=\tilde{{\bf G}}(\rho^{(1)},\tilde{{\bf x}})-M_{x3}\dot{\rho}^{(1)}.
\end{equation}
The final simulation model is given by
\begin{align}
\dot{\tilde{{\bf x}}} & =M_{x4}^{-1}\bigl(\tilde{{\bf G}}(\rho^{(1)},\tilde{{\bf x}})-M_{x3}\dot{\rho}^{(1)}\bigr)\\
\dot{\rho}^{(1)} & =\frac{{d}}{{d}t}\rho^{(1)}
\end{align}
where $\rho^{(1)}$ is specified by the user. If constant pressure is assumed, $\dot{\rho}^{(1)}=0$ and the formulation simplifies.
The primary drawback of this simulation technique is that source fluxes are treated as unconstrained injections. Thus, linepack depletion cannot be explored properly in contingency situations, since the system has an infinite flux reservoir at its disposal. 

\subsection{Technique 2: Finite Flux Reservoir with Upper Bound Modeled via Sigmoid Function}
As a realistic modification to the infinite flux reservoir model, we may hypothesize the existence of a slack node with two nonideal assumptions: (i) its flux output has a finite upper limit, and (ii) its density of injection decreases in value as flux injection saturates. The first assumption turns an infinite flux reservoir into a finite one, which enables our implicit ODE model to consider system-wide flux imbalance { while retaining regularity}. The second assumption stems from a more realistic consideration of satisfying the operating curves of centrifugal compressors which are widely used in the natural gas industry \citep{boyce:1993principles}. As the mass flux reaches its rated flow at the rated density (pressure) value, further increase of the mass flux requires a decline of the density to avoid the ``choke'' \citep{boyce:1993principles} or ``stonewall'' \citep{davis:1972simulation} phenomenon with potentially damaging vibrations. 

To create this nonideal slack node, we introduce a fictitious state variable $z$ which parameterizes the mass flux flowing from a slack source. In particular, we consider a situation where the reservoir at a source has an output flow limit; we may codify this limit with a sigmoid function. Instead of imposing constant input $d$, we parameterize it by the fictitious variable $z$, such that the value of the input $d$ is flexible, but within a certain limit. Replacing the specified injection term $d$ with the injection function ${\overline \phi}_{m} S_1(z)$, the flux injection at the source can be written as
\begin{equation}\label{eq: inj_Sig}
\phi_{0}={\overline \phi}_{m} S_1(z),
\end{equation}
where $S_1(z)$ is the sigmoid function given by
\begin{equation}
S_1(z)=\frac{e^{z}}{1+e^{z}}
\end{equation}
and the constant ${\overline \phi}_{m}$ represents the maximum upper limit of flux which the source can produce; this limit is approached as $z\rightarrow\infty$. 

As the flux flow saturates, according to the typical operating curve \citep{boyce:1993principles}, we parameterize the density of the source node as a monotonically decreasing function of flux. To do so, we write 
\begin{align}\label{eq: dens_Sig}
\rho=\overline{\rho}_n S_2(\phi_0),
\end{align}
where $\overline{\rho}_{n}$ is some nominal density of the node, and $S_2(\phi_0)$ is a sigmoid function flipped about the y-axis:
\begin{align}
    S_2(\phi_0)&=\frac{e^{\gamma(\overline{\phi}_{M}-\phi_{0})}}{1+e^{\gamma(\overline{\phi}_{M}-\phi_{0})}}.
\end{align}
The constant $\overline{\phi}_{M}$ represents the flux value at which the density reaches half its nominal value, and the constant $\gamma>0$ controls the speed at which the nodal density decreases as the flux injection approaches saturation\footnote{The numerical value of $\gamma$ can be inferred from the real operating curve or measurement data.}. The sigmoid function could also be replaced with a fitted polynomial of arbitrary degree. Using the chain rule $(\frac{df}{dt}=\frac{df}{dx}\frac{dx}{dt})$, the flux and density function time derivatives are 
\begin{align}
\dot{\phi}_{0} & =\underbrace{\overline{\phi}_{m}\left(S_{1}(z)-S_{1}^{2}(z)\right)}_{h_{1}(z)}\dot{z}\label{eq: sig_diff1}\\
\dot{\rho} & =\underbrace{\overline{\rho}\gamma\left(S_{2}^{2}(\phi_{0})-S_{2}(\phi_{0})\right)}_{h_{2}(\phi_0)}\dot{\phi}_{0}.\label{eq: sig_diff2}
\end{align}
As a clarification, the slack node density (pressure) $\rho$ is \textit{not} a dynamic state variable, but knowledge of its derivative is still essential for simulating the network. We again assume, WLOG, that the first node in $\Sigma_c$ is treated as a slack node, with constrained flux state $\phi_0$ and algebraic density $\rho$. We also reorder system $\bf G$ such that the conservation law equation associated with the slack node is altered to (\ref{eq: sig_diff1}) and placed at the end of the equation vector. Thus, by borrowing the formulation from (\ref{eq: parsed_sys}), the updated system may be described according to
\begin{align}\label{eq: Sig_initial_statement}
\left[\begin{array}{c|c|c}
M_{x3} & M_{x4} & {\bf 0}\\
\hline 0 & -{\bf e}_{k} & h_1(z)
\end{array}\right]\left[\begin{array}{c}
\dot{\rho}\\
\hline \dot{\tilde{{\bf x}}}\\
\hline \dot{z}
\end{array}\right]=\left[\begin{array}{c}
\tilde{{\bf G}}(\rho,\tilde{{\bf x}})\\
\hline 0
\end{array}\right]
\end{align}
where ${\bf e}_k$ is a row vector of zeros with a single 1 at index $k$, i.e. the index of state variable $\phi_0$. Since $M_{x3}{\dot \rho} = M_{x3} h_2(\phi_0)\dot\phi_0$, then (\ref{eq: Sig_initial_statement}) may be simplified to
\begin{align}\label{eq: Sig_model_mats}
\underbrace{\left[\begin{array}{c|c}
M_{x5} & {\bf 0}\\
\hline -{\bf e}_{k} & h_{1}(z)
\end{array}\right]}_{M_{s}}\left[\begin{array}{c}
\dot{\tilde{{\bf x}}}\\
\hline \dot{z}
\end{array}\right] & =\left[\begin{array}{c}
\tilde{{\bf G}}(\rho,\tilde{{\bf x}})\\
\hline 0
\end{array}\right].
\end{align}
If slack flux $\phi_0$ is the final state variable in $\tilde {\bf x}$, then $M_{x5}$ is
\begin{align}
M_{x5}=\left[M_{x4}^{(1)}\mid  M_{x4}^{(2)}\mid \cdots \mid M_{x4}^{( k)}+h_{2}(\phi_{0})M_{x3} \mid \cdots \mid M_{x4}^{(n)} \right],
\end{align}
where $M_{x4}^{(i)}$ is the $i^{\rm th}$ column of matrix $M_{x4}$. By Theorem \ref{theorem: delete_conservation}, matrix $M_{x4}$ is nonsingular. The vector $h_{2}(\phi_{0})M_{x3}$ has its only nonzero (and negative definite) entry at the index of the slack node density state variable. When viewed as a perturbation of $M_{x4}$, for all practical purposes, $M_{x5}$ will also be a nonsingular matrix. To simulate the system (\ref{eq: Sig_model_mats}), $M_s$ must be inverted. By inspection, the inverse of $M_s$ is given by
\begin{align}\label{eq: mat_inv_sig}
M_s^{-1} & =\left[\begin{array}{c|c}
M_{x5}^{-1} & {\bf 0}\\
\hline\frac{M_{x5}^{-1}(k,:)}{h_1(z)} & \frac{1}{h_1(z)}
\end{array}\right],
\end{align}
where $M_{x5}^{-1}(k,:)$ refers to the $k^{\rm th}$ row of $M_{x5}^{-1}$.
\begin{theorem}
    Matrix $M_s$ of (\ref{eq: Sig_model_mats}) is nonsingular if $|z|<\infty$.
    \begin{proof}
    Since matrix $M_{x5}$ is full rank, square matrix $M_s$ will clearly have full row rank as long as $h_1(z)\ne 0$, since no linear combination of the above rows could create the bottom row. Assuming ${\overline \phi}_{m} \ne 0$, $h_1(z)$ of (\ref{eq: sig_diff1}) is equal to 0 only when $S_1(z)=0$ or $S_1(z)=1$, which only occur when $z=\pm \infty$.
    \end{proof}
\end{theorem}
When a source flux injection is modeled by a sigmoid function, the fictitious state variable $z$ will tend to ``blow up" when the upper flow limit ${\overline \phi}_{m}$ is approached. This is problematic numerically, rather than physically, so a helpful workaround is to artificially constrain the growth of $z$. 
{This may be accomplished through the application of a constraint function $\zeta(z)=1+e^{z^{2}-r^{2}}$ to $h_1(z)$, where $r$ is a constant scalar that is chosen by the user. In our simulations, we let $r=10$. Applying $\zeta(z)$ to $\dot{z}$ in (\ref{eq: sig_diff1}) we have,
\begin{align}\label{eq: constraint_func}
\dot{z}=\dot{\phi}_{0} h_{1}(z)^{-1} \zeta(z)^{-1},
\end{align}%
Since $e^{z^2}$ increases much faster than $e^z$, the $\zeta(z)^{-1}$ term essentially { forces $\dot{z}$ to decline to nearly zero very rapidly}. The matrix inversion expression of (\ref{eq: mat_inv_sig}) can be updated by simply replacing $h_1(z) \leftarrow h_1(z)\cdot \zeta(z)$}.

\subsubsection*{Simulation Model Statement}To state the model compactly, we augment the state variable vector $\tilde{\bf x}$ from (\ref{eq: xtilde}) by adding $z$: $\tilde{\bf x}_a = \tilde{\bf x}^\frown z$. We also augment ${\tilde{\bf G}}$ by adding a zero to the bottom row to form augmented vector ${\tilde{\bf G}}_a$:
\begin{align}
    \dot{\tilde{{\bf x}}}_a & =M_{s}^{-1}\tilde{{\bf G}}_a(\rho,\tilde{{\bf x}}_a)\\
    \rho &= \overline{\rho}_n S_2(\phi_0).
\end{align}
This model may be used to investigate how a system responds when source mass flux flow limits are reached.

\subsection{Technique 3: Constant Flux Sources}
While Technique 1 assumed infinitely variable mass flux sources and technique 2 assumed constrained mass flux sources, technique 3 assumes constant (or specified) mass flux sources at all nodes. This is ultimately accomplished by converting a density state variable into an algebraic variable. According to Theorem \ref{theorem: delete_pressure}, the deletion of a density state will render a mass matrix with full column rank. While Theorem \ref{theorem: delete_conservation} showed that we may delete a mass flux conservation equation to ensure full row rank, the following theorem shows that we may instead delete a momentum conservation equation from (\ref{eq: de3}) to ensure nonsingularity of $M_{\bf x}$.

\begin{theorem}\label{theorem: delete_conservation_momentum}
    Consider system $\Sigma_c$. The elimination of a pressure state and a flux momentum equation in (\ref{eq: de3}) from $M_{\bf x}$ will yield a square, full rank matrix $M_{\bf x}'''$.
    \begin{proof}
    We borrow matrix ${\overline M}_{{\bf x}}''$ from Theorem \ref{theorem: delete_conservation}, but we alter its definition by the elimination of one row (row $i$) from the bottom subsection of $M_{\bf x}'$ instead of the top:
    \begin{align}
           {\overline M}_{{\bf x}}'' = \left[\begin{array}{cc}
            { K}_{0}      & { K}_{L}\\
            \Gamma_{2}' & \Gamma_{2}'
            \end{array}\right].\label{eq: M_bpp2}
    \end{align}
    We now assume corresponding line $i$ connects nodes $j$ and $k$.  Our goal is to show that matrix (\ref{eq: M_bpp2}) has full row rank. To do so, we remove two columns from (\ref{eq: M_bpp2}): column $i$ and column $i+m$. Each of these columns will have a single non-zero entry, at indices $j$ and $k$ respectively, so the row rank of the matrix necessarily drops by 2. By eliminating these columns, we are effectively eliminating line $i$ from the incidence matrix and our system $\Sigma_c$ is no longer fully connected. We now eliminate rows $j$ and $k$. This effectively eliminates nodes $j$ and $k$. By performing these operations, we are left with matrices which we define to be called ${\hat K}_0$, ${\hat K}_L$, and ${\hat \Gamma}_2'$. Implicitly contained in ${\hat K}_0$ and ${\hat K}_L$ are two tree-structured subgraphs which have both been reduced. By defining
    \begin{align}
           {\widehat{\overline M}}_{{\bf x}}'' = \left[\begin{array}{cc}
            {\hat K}_{0}      & {\hat K}_{L}\\
            \hat\Gamma_{2}' & \hat\Gamma_{2}'
            \end{array}\right],\label{eq: M_bppH}
    \end{align}
    we employ the same tactics used in the Theorem \ref{theorem: delete_conservation} proof:
        \begin{subequations}\label{eq: Mxbhpp} 
        \begin{align}
            {\rm det}\Bigl({\widehat{\overline M}}_{{\bf x}}''\Bigr) & ={\rm det}\bigl({\hat K}_{0}-{\hat K}_{L}\hat\Gamma_{2}^{-1}\hat\Gamma_{2}\bigr){\rm det}\bigl(\hat\Gamma_{2}\bigr)\\
             & ={\rm det}\bigl({\hat K}_{0}-{\hat K}_{L}\bigr)\sigma.
        \end{align}
        \end{subequations}
        where $\sigma\ne 0$ since $\hat\Gamma_{2}$ is a diagonal matrix. Matrix $\hat K_{0}-\hat K_{L}$ is in fact a nonsingular matrix, because it represents the block diagonal concatenation of two reduced (and scaled) tree incidence matrices which are in themselves both nonsingular. Since the determinant in (\ref{eq: Mxbhpp}) is nonzero, then the following is implied:
        \begin{align}
        {\rm rank}({\overline M}_{{\bf x}}'') &= {\rm rank}({\widehat{\overline M}}_{{\bf x}}'')+2\\
        & = n+m-1.
        \end{align}
        because full row rank of a square matrix implies full column rank. Thus, (\ref{eq: M_bpp2}) is a full rank matrix. By direct extension, matrix $M'''_{{\bf x}}$, which is defined by
        \begin{align}
           M'''_{{\bf x}} = \left[\begin{array}{ccc}
            {\bf 0} & K_{0} & K_{L}\\
            \Gamma_{1}' & {\bf 0} & {\bf 0}\\
            {\bf 0} & \Gamma_{2}' & \Gamma_{2}'
            \end{array}\right],
        \end{align}
        is also full rank.
    \end{proof}
\end{theorem}
We now leverage the results of Theorem \ref{theorem: delete_conservation_momentum} in order to define a simulation technique which allows for flux injections to be simultaneously specified at all nodes. To do so, we define a balancing node.
\begin{definition}
    A \textbf{balancing node} is a node whose flux injection is specified and whose nodal density $\rho$ is transformed from a state variable into an algebraic variable by solving a momentum balance equation.
\end{definition}
The balancing node's density variable thus becomes an algebraic variable whose value can be computed analytically. The decision of which node is chosen to be this balancing node is an important point which shall be considered later in this subsection. For now, we consider some balancing node which is connected to the rest of the tree through a single line\footnote{Any node in the network can be selected as the balancing node, so long as the dynamics of an interconnecting line can be described via (\ref{eq: momentum_int_2}).}. The momentum equation associated with its interconnection line is defined as
\begin{align}
\!\!\frac{l}{2}(\dot{\phi}_{l}\!+\!\dot{\phi}_{0})+a^{2}\left(\rho_{l}\!-\!\rho_{0}\right)&=\frac{-l\lambda}{4D}\left(\phi_{l}\!+\!\phi_{0}\right)\!\frac{|\phi_{l}\!+\!\phi_{0}|}{\rho_{l}\!+\!\rho_{0}},\label{eq: momentum_int_2}
\end{align}
where we have neglected nodal indices for notational simplicity. Variables ${\phi}_{0}$ and $\rho_{0}$ refer to the flux flow and density at the balancing node, respectively. At this balancing node, we assume flux is exogenously specified. WLOG, we further assume constant flux such that $\dot{\phi}_{0}=0$. Thus, we may form a quadratic equation in $\rho_0$:
\begin{align}
0 & = \left[-\frac{2a^{2}}{l}\right]\rho_{0}^{2}+\left[\dot{\phi}_{l}\right]\rho_{0}+\nonumber\\
&\left[\frac{\lambda}{2D}\left(\phi_{l}+\phi_{0}\right)|\phi_{l}+\phi_{0}|+\dot{\phi}_{l}\rho_{l}+\frac{2a^{2}}{l}\rho_{l}^{2}\right].\label{eq: qf}
\end{align}
The value of $\rho_{0}$ is a function of state variables $\phi_0$, $\phi_l$ and $\rho_l$ along with state variable derivative $\dot{\phi}_{l}$. Assuming all of these values are known numerically, $\rho_{0}$ can be computed analytically via the quadratic formula. We thus write $\rho_{0}$ as a function $g$ of state variable vector $\bf x$ and derivative $\dot{\phi}_{l}$:
\begin{equation}
    \rho_{0} = g({\bf x},\dot{\phi}_{l}).
\end{equation}
To compute $\rho_{0}$, we must know ${\bf x}$ and $\dot{\phi}_{l}$ numerically. The state variable values of $\phi_0$, $\phi_l$ and $\rho_l$ are certain to be known (at each numerical time step). In the following theorem, we show that $\dot{\phi}_{l}$ can be numerically computed without numerical knowledge of $\rho_{0}$ or its time derivative.
\begin{theorem}\label{Theorem flux_deriv}
The numerical value of the flux derivative $\dot{\phi}_{l}$ at the far end of the line attached to a balancing node may be computed without knowledge of the density at the balancing node $\rho_{0}$ or its derivative ${\dot \rho}_{0}$.
\begin{proof}
Consider some system of equations given by
\begin{align}
K_{l}\dot{\boldsymbol{\phi}}_{l}+K_{0}\dot{\boldsymbol{\phi}}_{0} & =\dot{\boldsymbol{d}}\\
[\Gamma_{2}(\dot{\boldsymbol{\phi}}_{l}+\dot{\boldsymbol{\phi}}_{0})]' & =\left[\Gamma_{3}\boldsymbol{\rho}-\Gamma_{4}{\bf f}\left(\boldsymbol{\phi}_{l},\boldsymbol{\phi}_{0},\Gamma_{5}\boldsymbol{\rho}\right)\right]',\label{eq: mom_p}
\end{align}
where the prime notation indicates the deletion of one momentum equation, i.e. (\ref{eq: momentum_int_2}). The mass matrix associated with this system exactly corresponds to (\ref{eq: M_bpp2}) which has been shown to be nonsingular. If the momentum equation associated with the balancing node's line is the equation which was deleted, then (\ref{eq: mom_p}) will not contain the density of the balancing node or its derivative. Therefore, all flux flow derivatives, including $\dot{\phi}_{l}$, may be solved for through matrix inversion of (\ref{eq: M_bpp2}), without knowledge of $\rho_{0}$ or ${\dot \rho}_{0}$
\end{proof}
\end{theorem}
With the results of Theorem \ref{Theorem flux_deriv}, we may assume there is some function $y$ which computes the necessary flux flow derivative value: $\dot{\phi}_{l}=y({\bf x})$. Thus, we may solve (\ref{eq: qf}) via the quadratic formula if we first substitute in $\dot{\phi}_{l}=y({\bf x})$. 

While the solution to (\ref{eq: qf}) is necessary, we shall show that the model associated with this simulation technique additionally depends on the density derivative ${\dot \rho}_{0}$ at the balancing node. The value of ${\dot \rho}_{0}$ may be computed via the chain rule:
\begin{subequations}
\begin{align}
\dot{\rho}_{0} & =\frac{d}{{d}t}g({{\bf x}},\dot{\phi}_{l})\label{eq: dp0_dgdt}\\
 & =\sum_{i}\frac{d}{{d}{{\bf x}}_{i}}g({{\bf x}},y({\bf x}))\dot{{{\bf x}}}_{i}.\label{eq: dp0_sum}
\end{align}
\end{subequations}
The vector $\dot{{{\bf x}}}_{i}$ will include flux flow derivatives as well as density derivatives. As proved in Theorem \ref{Theorem flux_deriv}, the flux flow derivatives can be computed without $\rho_0$ or ${\dot \rho}_0$. The density derivatives are considered in the following theorem.
\begin{theorem}\label{eq: theorem_density_derivs}
Once a balancing node has been selected, the density state derivatives $\dot{\boldsymbol{\rho}}$ may be uniquely computed.
\begin{proof}
When the expression $\Gamma_{1}\dot{\boldsymbol{\rho}} =\boldsymbol{\phi}_{0}-\boldsymbol{\phi}_{l}$ from (\ref{eq: de2}) has been altered such that a density state derivative is removed, $\Gamma_{1}$ loses a column and is transformed into $\Gamma_1'$ from (\ref{eq: Mxp}). Because $\Gamma_1'$ has been shown to be a nonsingular matrix, the system in (\ref{eq: de2}) may be solved by specifying a single density derivative (i.e. the one that was removed) if both $\boldsymbol{\phi}_{0}$ and $\boldsymbol{\phi}_{l}$ are known.

Alternatively, instead of specifying a density derivative state to solve (\ref{eq: de2}), a new linear equation may be introduced which specifies the relationship of the ``eliminated" state with the remaining states. Since (\ref{eq: dp0_sum}) represents such a linear relationship, and since this relationship is physically independent from the processes used to model the relationships codified by (\ref{eq: de2}), then a unique solution for the density derivative states may be solved for.
\end{proof}
\end{theorem}

We thus seek to use (\ref{eq: dp0_sum}) in order to add a new linearly independent row to $\Gamma_1$. To do so, we restate (\ref{eq: dp0_sum}) as
\begin{align}
\underbrace{\left[\begin{array}{cccc}
\!1 & -a_{1} & \cdots & -a_{n}\!\end{array}\right]}_{\bf v}\dot{\boldsymbol{\rho}}=b,
\end{align}
where $\dot{\boldsymbol{\rho}}$ is the unknown vector, and known $a_i$, $b$ are given by
\begin{align}
a_{i}&=\frac{d}{d{\bf x}_{i}}g({\bf x},y({\bf x})),\;\,i\in\{1...n\}\\
b    &= \sum_{i\notin\{1...n\}}\!\frac{d}{d{\bf x}_{i}}g({\bf x},y({\bf x}))\dot{{\bf x}}_{i}.
\end{align}
Thus, we may build augmented matrix $\Gamma_{1a}$ by appropriately concatenating matrix $\Gamma_1$ and vector ${\bf v}$: $\Gamma_{1a} = \Gamma_1^\frown{\bf v}$. Finally,
\begin{align}\label{eq: rho_dot_sol}
    \dot{\boldsymbol{\rho}}=\Gamma_{1a}^{-1}\left[\!\begin{array}{c}
    \boldsymbol{\phi}_{0}-\boldsymbol{\phi}_{L}\\
    b
    \end{array}\!\!\right].
\end{align}

\subsubsection*{Simulation Model Statement}
In order to explicitly state the model associated with this simulation technique, we restate (\ref{eq: parsed_sys}), but we reorder the system $\bf G$ into ${\bf G}^r$ such that the first equation ${\bf G}_1^r$ is the momentum equation associated with the line attached to the balancing node:
\begin{equation}\label{eq: parsed_sys_reorder}
\left[\begin{array}{c|c}
M^r_{x1} & M^r_{x2}\\
\hline M^r_{x3} & M^r_{x4}
\end{array}\right]\left[\begin{array}{c}
\dot{\rho}^{(1)}\\
\hline \dot{\tilde{{\bf x}}}
\end{array}\right]=\left[\begin{array}{c}
{\bf G}^r_{1}(\rho^{(1)},\tilde{{\bf x}})\\
\hline \tilde{{\bf G}}^r(\rho^{(1)},\tilde{{\bf x}})
\end{array}\right],
\end{equation}
where the first density variable is also associated with the balancing node. This system may be solved according to
\begin{align}
\dot{\tilde{{\bf x}}} &=M_{x4}^{r-1}\bigl(\tilde{{\bf G}}^r(\rho^{(1)},\tilde{{\bf x}})-M^r_{x3}\dot{\rho}^{(1)}\bigr)\\
\rho^{(1)} &= g(\tilde{\bf x},y(\tilde{\bf x}))\\
\dot{\rho}^{(1)} & =\frac{d}{{d}t}g({\tilde{\bf x}},y(\tilde{\bf x})),
\end{align}
where the first equation from (\ref{eq: parsed_sys_reorder}) is used to build function $g$, the results of Theorem \ref{Theorem flux_deriv} are used to build function $y$, and the results of Theorem \ref{eq: theorem_density_derivs} and (\ref{eq: rho_dot_sol}) may be used for taking the derivative of the density. In a large network, it may be numerically expedient to write (\ref{eq: qf}) as $0=q(\rho_0,{\bf x},{\dot \phi}_l)=q(\rho_0,{\bf x},y({\bf x}))=Q(\rho_0,{\bf x})$. In noting that
\begin{align}\label{eq: implicit_derivative}
0 & =\frac{dQ}{d\rho_0}\dot{\rho}_0+\sum_{i}\frac{dQ}{d{\bf x}_{i}}\frac{d{\bf x}_{i}}{dt},
\end{align}
the time derivative $\dot{\rho}_0$ may be solved for by rearranging (\ref{eq: implicit_derivative}).

\subsubsection*{Selecting the Balancing Node}
In a system where all flux injections are specified, it may not be obvious which node should be selected as the balancing node. The physical characteristics of this selection are important to consider, though. Since the balancing node's density variable will transform into an algebraic variable, it will have the obligation of reacting instantaneously to any local imbalance or change in the system. For this reason, the node selected as the balancing node should generally be the node closest to the location of any system disturbance (e.g. loss of compressor, increase in load, loss of supply, leak in line). This choice, though, should be thought of as a \textit{dynamic} choice, because if there is a subsequent perturbation at the source, the source should then be reassigned as the balancing node. In this way, any time the system is perturbed, the balancing node should be reassigned nearest to the location of the perturbation.




\section{Test Results}\label{Test Results}
\begin{figure}[ht!]
\centering
\includegraphics[width=0.6\textwidth]{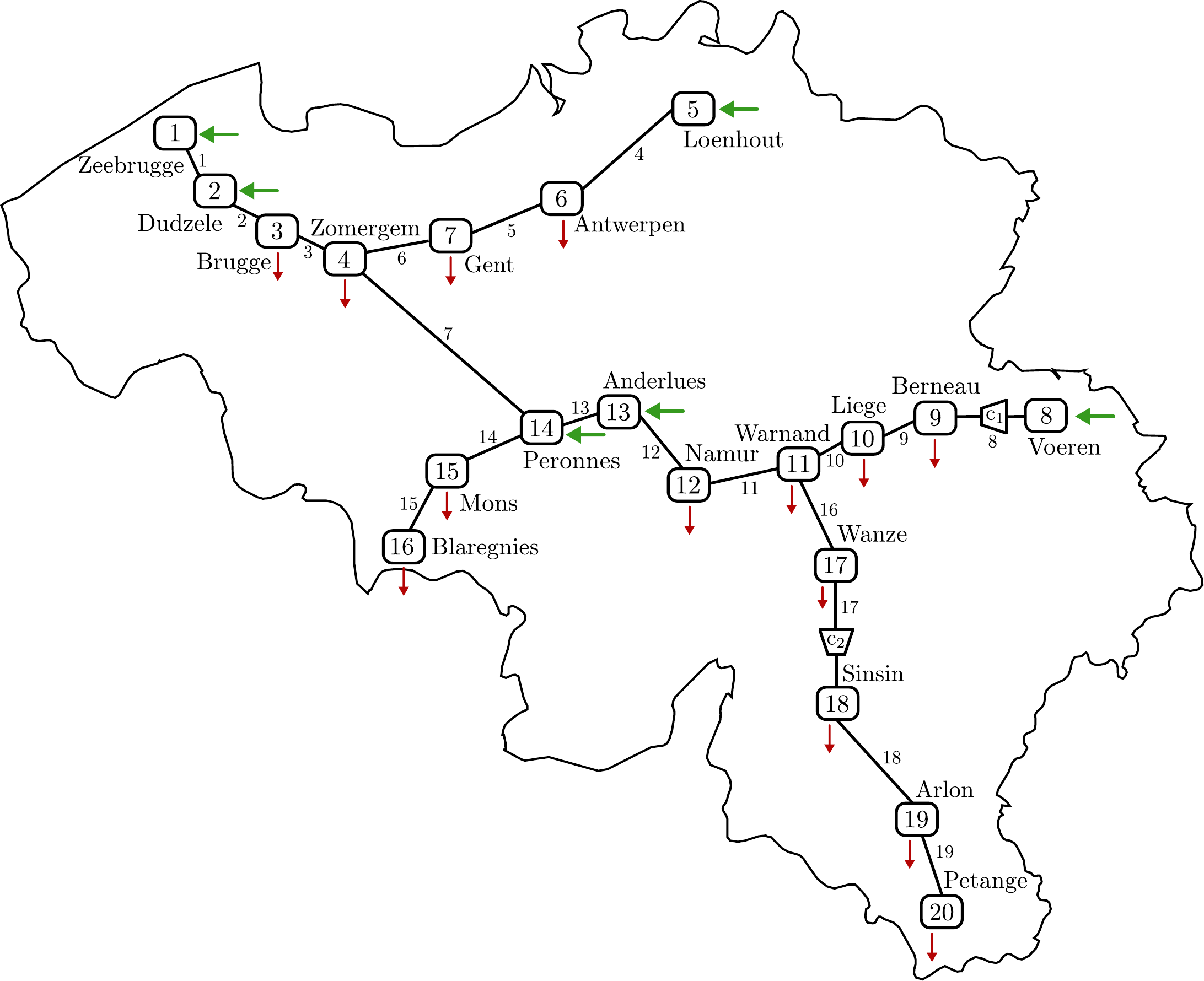}
\caption{The radial Belgium system has 20 nodes, 19 lines, and 2 active compressor stations. All parallel lines from the original model presented in~\citep{Wolf:2000} have been combined.\label{fig: Belgium}}
\end{figure}

In this section, we compare the previously derived simulation techniques via tests on the 20 node Belgium network; this system was reconstructed based on the model presented in~\citep{Wolf:2000}, with some alterations. As shown in Fig. \ref{fig: Belgium}, this radial network has 19 lines and two active compressors ($\rm c_1$ and $\rm c_2$). The longest line in the network is 98 km long, but all lines are finitely discretized into $l=5$ km sections (including the interconnection to the balancing node). In total, this discretization yields approximately 300 state variables in each simulation. Simulation code and network data are posted online for open source access\footnote{github.com/SamChevalier/Natural-Gas-Simulations-TCNS}. All simulations were performed using MATLAB's ode23tb function on a Dell XPS laptop, equipped with an Intel i5 CPU @ 2.30GHz and 8 GB of RAM. Standard MATLAB ODE simulation tolerances and a time step resolution of $\Delta t = 0.5$s were employed. The computational time for each scenario is less than a minute.

\subsection{Test 1: Sharp Load Increase at Node 16}
\begin{figure}[ht!]
\centering
\includegraphics[width=0.95\columnwidth]{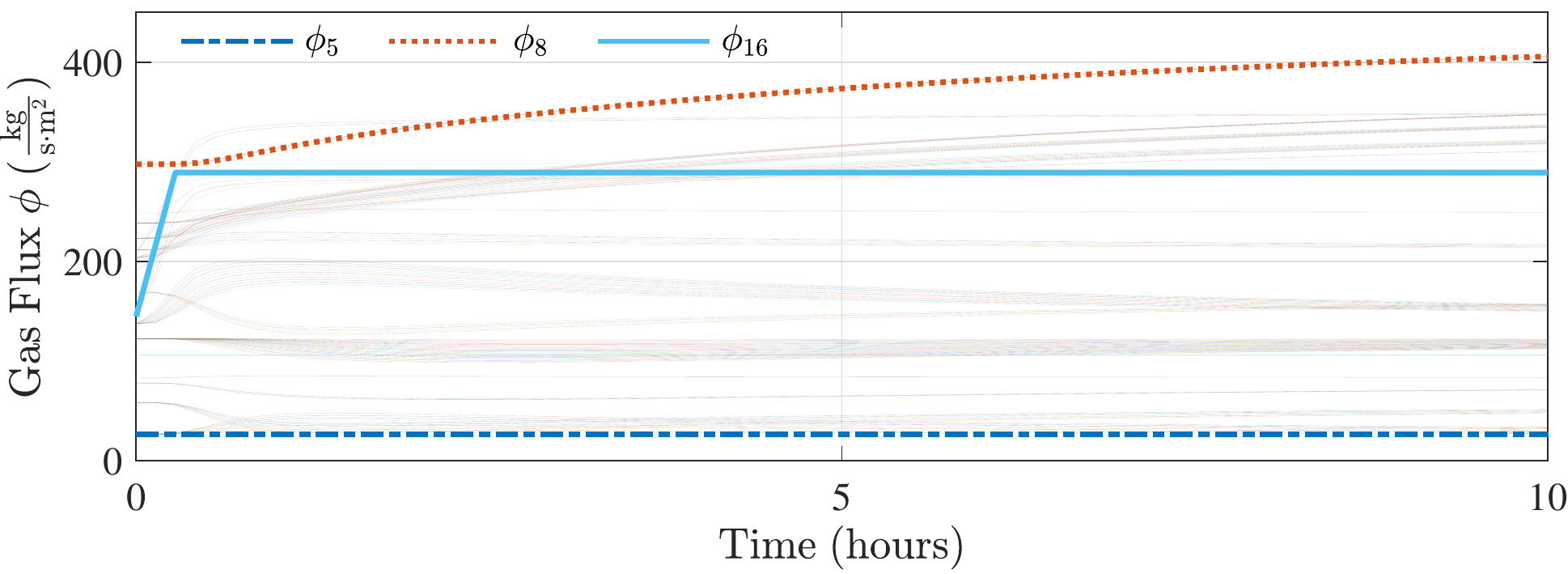}
\caption{Shown is the system-wide flux response for a ramp increase in load at node 16. The load ramps over the first 1000s of simulation. These results are associated with the infinite flux reservoir simulation.\label{fig: R1_FluxT1}}
\end{figure}

In this test, the load at node 16 was doubled\footnote{Increasing load by 90 ${\rm kg}/{\rm s}$ increased total system load by 34\%.} in value over the course of 1000 seconds. 
This was accomplished by setting ${\dot {\boldsymbol d}}_{16}$ from (\ref{eq: de1}) to $-0.09$ ${\rm kg}/{\rm s}^2$ from time $t=0$ to $t=1000$, as shown by $\phi_{16}$ in Fig. \ref{fig: R1_FluxT1}. This smooth load growth prevented the excitation of any high frequency transients, which the given models aren't suitable for capturing. The system was further simulated for another 60 hours (or until bifurcation) with all three alternative modeling techniques. The following subsections describe the results of each simulation in more detail.

\begin{figure}[ht!]
\centering
\includegraphics[width=0.95\columnwidth]{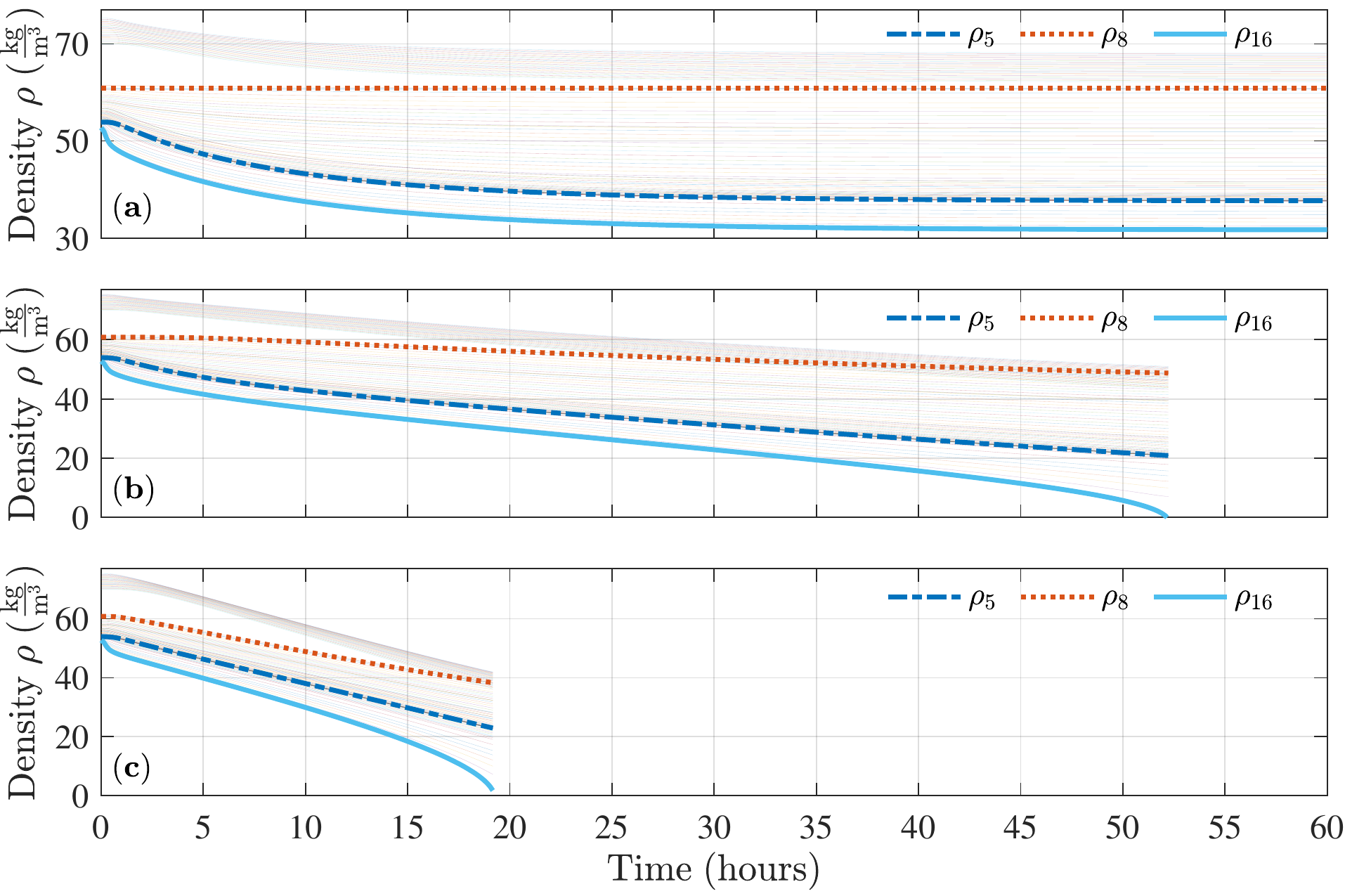}
\caption{Shown is the system-wide density response for a ramp increase in load at node 16. Panel $({\bf a})$ shows the infinite flux reservoir response; panel $({\bf b})$ shows the bounded finite flux reservoir response; panel $({\bf c})$ shows the constant flux source response.\label{fig: R1_DensityT1}}
\end{figure}

\subsubsection{Technique 1: Infinite Flux Reservoir}
In this trial, source node 8 was chosen as the constant density slack node. The density response for all system nodes (real and virtual) is shown in panel $({\bf a})$ of Fig. \ref{fig: R1_DensityT1} across the full 60 hours of simulation. As can be seen, after approximately 50 hours, the system converged to a new steady state operating equilibrium. The slack node was able to fully compensate for the load increase, and the flux imbalance in the system was ultimately driven to 0. The density at node 8 (the slack node) remained constant despite a precipitous fall in pressure at other nodes in the network. The pressure at node 5, which is also a source node, dropped considerably, because this source was modeled as a constant flux injector. Due to the limitless flux injection at the slack node, system linepack was never be fully depleted with this simulation model.

\subsubsection{Technique 2: Finite Flux Reservoir with Upper Bound Modeled via Sigmoid Function}
In this trial, source node 8 was converted from a constant density node to a node whose flux injection was constrained by (\ref{eq: inj_Sig}) and whose density of injection was governed via (\ref{eq: dens_Sig}). While the mass flow consumed by the load increased by 90 ${\rm kg}/{\rm s}$, ${\overline \phi}_m$ was chosen such that the source could only provide an additional 60 ${\rm kg}/{\rm s}$ of mass flow, inducing a sustained mass flow imbalance at -30 ${\rm kg}/{\rm s}$. Additionally, $\gamma=0.1$ was selected. The densities for all system nodes (real and virtual) is shown in panel $({\bf b})$ of Fig. \ref{fig: R1_DensityT1} across the full 60 hours of simulation. 

At hour 52, the linepack of a pipeline in the system was completely depleted since the gas density (pressure) at node 16 reached zero. It was caused by the constant imbalance of the mass flow rate in the system due to the overloading at node 16 beyond the injection capability of source node 8.
As the flux at node 8 saturated, the linepack in the system began to deplete, and the system could no longer support the loads. When approaching node 8's maximum injection capability, the designated density (pressure) at node 8 could not be sustained anymore. Thus, node 8 experienced a continuous density drop as well, according to our model, which further contributed to the density drop at other nodes. 
The whole process is in significant contrast with the Infinite Flux Reservoir model results, shown in panel $({\bf a})$, where the flux was finally re-balanced and the system stabilized to a new equilibrium point. When simulated using this model, the survival time of the system is clearly finite. This comparison validates that our proposed Finite Flux Reservoir model can capture the slow dynamics of the linepack depletion for severe contingency situations, and it can provide an estimation of the system survival time when system sources saturate.

\subsubsection{Technique 3: Constant Flux Source}
This model is more restrictive than the previous two, since all the source node flux inputs are rigid without any flexibility. Thus, we expected that the system could survive for an even shorter period of time. In this trial, source node 8 was selected as the balancing node\footnote{This selection was made for comparison purposes. As previously noted, node 16 is actually the most suitable choice for balancing node, since this is the location of the disturbance.} Its density state was thus converted into an algebraic variable and its flux injection was held fixed. The density response for all system nodes (real and virtual) is shown in panel $({\bf c})$ of Fig. \ref{fig: R1_DensityT1} across the full 60 hours of simulation. The linepack of the system was completely depleted since the density of node 16 reached zero after only 19 hours, though. As can be seen, the pressure in the system collapse much faster than in the previous simulation. Nodal densities drop dramatically due to the severe flux imbalance in the network.

\subsection{Test 2: Partial Loss of Compressor at Node 17 during Load Increase Period}

\begin{figure}[ht!]
\centering
\includegraphics[width=0.95\columnwidth]{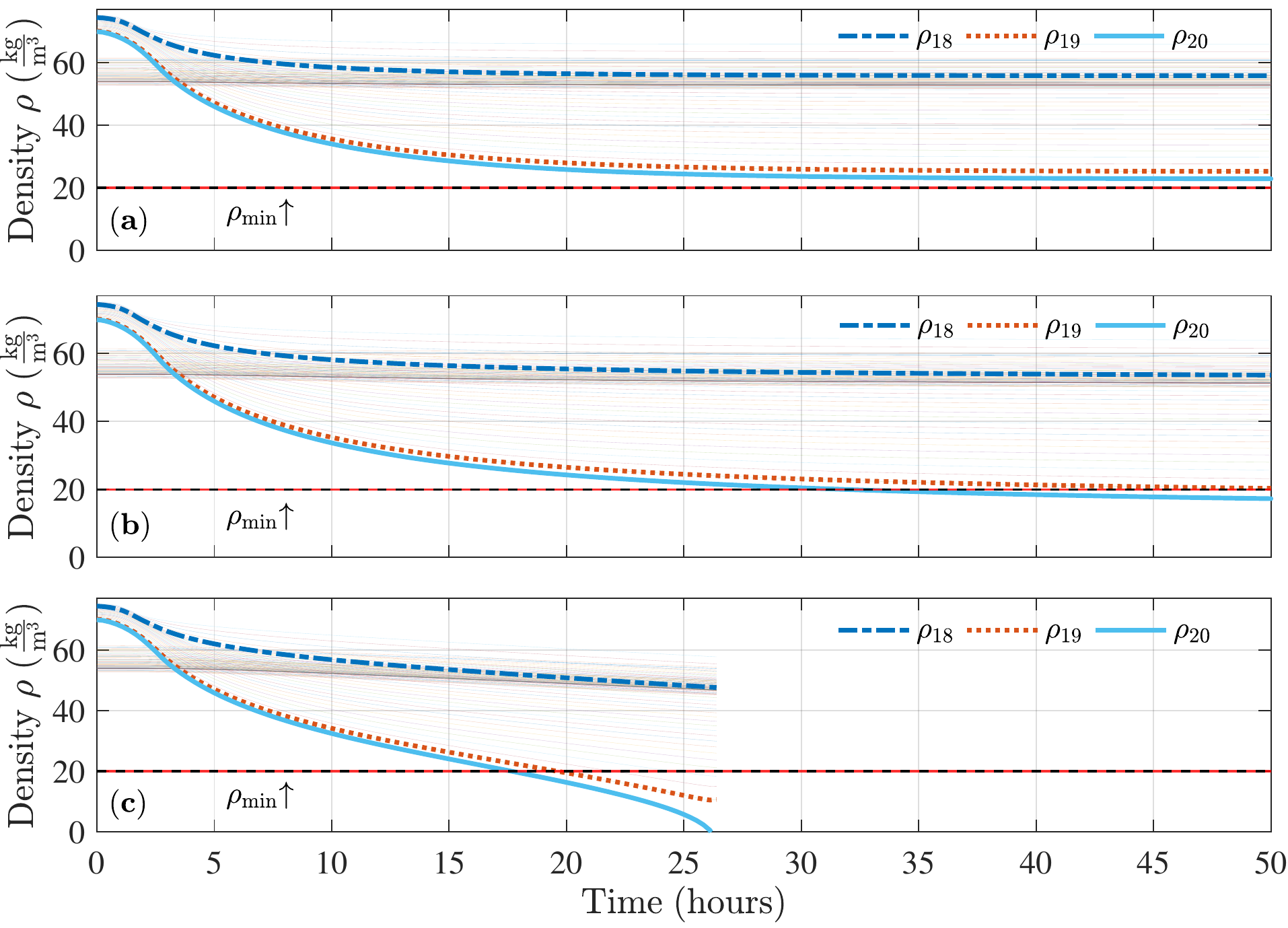}
\caption{Shown is the system-wide density response for a ramp increase in load at nodes 18, 19, and 20, along with the partial shutdown of a compressor at node 17. Panel $({\bf a})$ shows the infinite flux reservoir response; panel $({\bf b})$ shows the bounded finite flux reservoir response; panel $({\bf c})$ shows the constant flux source response. The minimum allowable density $\rho_{\rm min}$ is also identified.\label{fig: R1_DensityT2}}
\end{figure}

\begin{figure}[ht!]
\centering
\includegraphics[width=0.95\columnwidth]{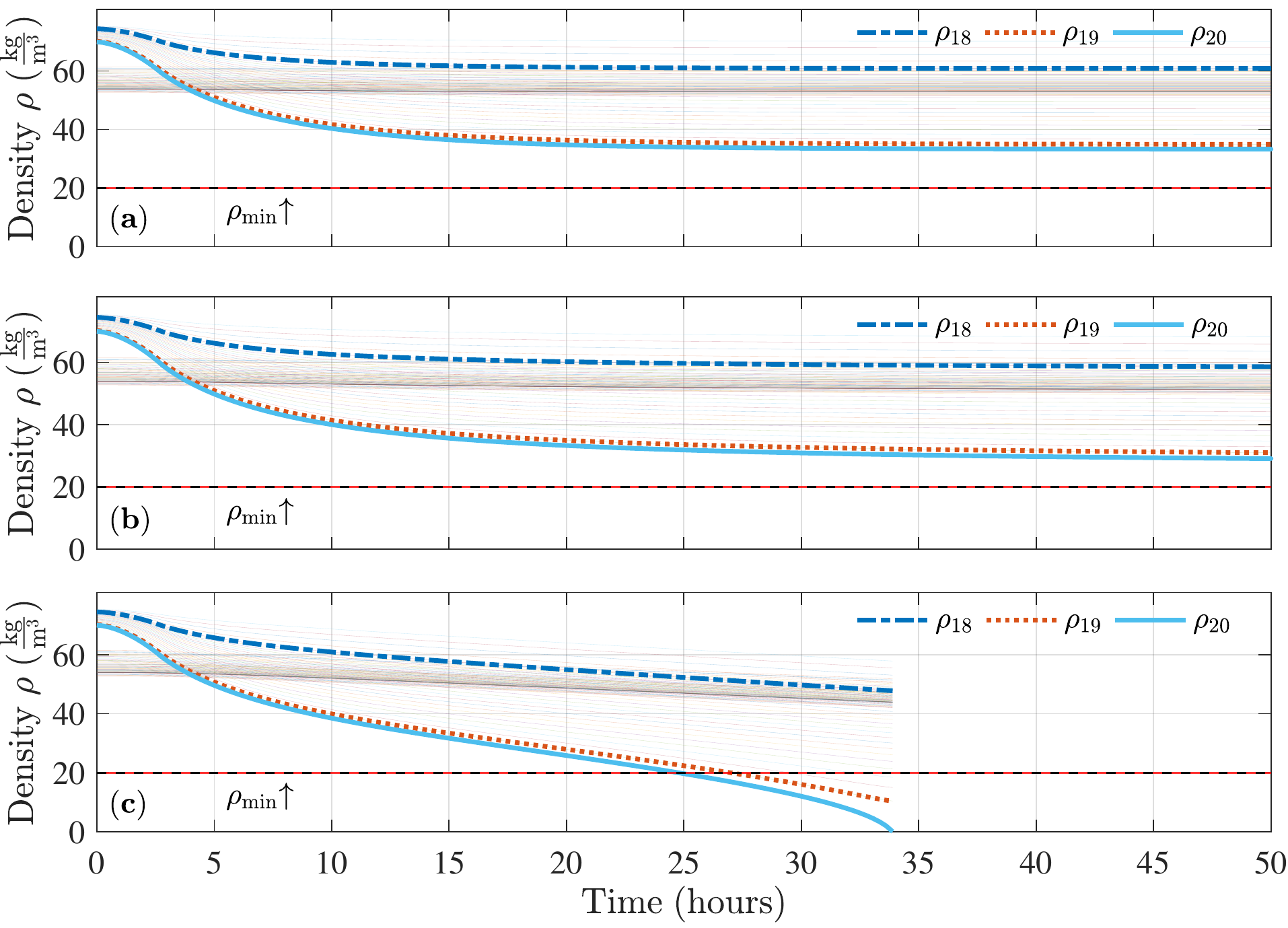}
\caption{Shown is the system-wide density response for a ramp increase in load at nodes 18, 19, and 20. In this case, \textit{no} partial shutdown of the node 17 compressor occurs. Panel $({\bf a})$ shows the infinite flux reservoir response; panel $({\bf b})$ shows the bounded finite flux reservoir response; panel $({\bf c})$ shows the constant flux source response. The minimum allowable density $\rho_{\rm min}$ is also identified.\label{fig: R1_DensityT2A}}
\end{figure}

In this test, we considered a contingency where a compressor was partially compromised during a load increasing period. We restricted the load increase such that it was within the source node injection capability of the second technique in order to avoid linepack depletion. We were interested in whether the system could survive in the post-fault period, and if it could survive, how the compressor loss affected the new steady state convergence. 

Specifically, the compressor at node 17 was intentionally compromised. Its original density amplification value of $30.28\%$ (i.e. $\alpha=1.3028$) was decreased by $25\%$ over the course of 2 hours using a smooth sigmoid function. Meanwhile, the loads at nodes 19 and 20 were doubled over this time period, and the load at node 18 was increased from 0 to 10 ${\rm kg}/{\rm s}$. These load changes were not beyond the maximum injection capability at source node 8 for the second technique. We further imposed that all loads had a minimum gas density (pressure) requirement of $\rho_{\rm min} = 20$. The system was simulated for fifty hours with the three alternative techniques; the density responses for all three simulations are shown in Fig. \ref{fig: R1_DensityT2}.

Panel $({\bf a})$ in Fig. \ref{fig: R1_DensityT2} corresponds to the infinite flux reservoir model simulation. The results show that the system converged to a stable equilibrium, with the lowest density in the system being $\rho \approx 23$, which was greater than $\rho_{\rm min}$. Therefore, the infinite flux reservoir model predicted that the system will survive the compressor fault during the load increasing period with a workable density condition.

Panel $({\bf b})$ corresponds to the bounded finite flux reservoir response. The system also converged to a stable equilibrium, but the lowest density was $\rho \approx 17$, which is less than $\rho_{\rm min}$. Thus, the finite flux reservoir model predicted that the system would survive the compressor fault, but some nodal densities were not workable.

Finally, panel $({\bf c})$ corresponds to the constant flux source simulation. Since all the source flux was fixed in this model, the overloading condition eventually drove the system to collapse. The partial loss of the compressor, though, accelerated this process for around 26 hours. 

These results can be contrasted to a situation where the loads at nodes 18-20 increase in the same manner, but the compressor at node 17 is \textit{not} compromised. Density responses are seen in Fig. \ref{fig: R1_DensityT2A}. The minimum density steady state responses of the system in panels ($\bf a$) and ($\bf b$) are both higher than their counterparts in Fig. \ref{fig: R1_DensityT2}, and both are well above $\rho_{\rm min}$. In panel ($\bf c$), the system does still collapse due to the flux imbalance, but the system's survival time is increased by approximately 8 hours due to the properly functioning compressor.

These tests demonstrate that our proposed techniques can be used to simulate various contingencies under different assumptions, and are capable of predicting the system survival time under the long term mass flux imbalance condition, which is not fully captured by the commonly used model.


\section{Conclusion}\label{Conclusion}
In this paper, three alternative { implicit ODE} models for studying dynamic linepack depletion in NGPNs are derived and discussed. The conventional model is shown to be degenerate under specified mass flux inputs for every node. Thus, it is not capable of characterizing linepack depletion in the case of a system-wide mass flux imbalance. The other two proposed implicit ODE models are shown to be {regular (non-degenerate)} under specified mass flux inputs. They are novel and present new opportunities for characterizing the survival time of NGPNs in the context of extreme contingencies. 

Using the proposed methods, we simulated and analyzed two specific potential contingencies: a sudden load increase, which could be caused by the emergency dispatch of gas-fired power plants or an unanticipated large scale leakage; and the partial loss of a compressor (during a load increase period), which could be caused by the malfunctioning of a compressor station or loss of power. In terms of evaluating linepack depletion and system survival time, the simulation results validated the anticipations of the proposed models: in both simulation experiments, the system had infinite survival time when a slack node was present in the system, but finite survival time when the slack node was fully converted to a constant flux source. Also as anticipated, the survival time for the constant flux system was shorter than the survival time for the system with the finite flux reservoir. Other contingencies, such as a line leak, system bifurcation, or sudden loss of a source, will be characterized in future publications. Additionally, this framework will be useful in characterizing the interdependency of different networks and the propagation of failures between electrical power systems and NGPNs.

\section*{Acknowledgements}\label{Acknowledgements}
The authors gratefully acknowledge support by the NSF grant CNS-1735463. We also thank Prof. Paul Hines at University of Vermont, Prof. Seth Blumsack at Penn State University, Prof. Eytan Modiano at MIT, and Dr. Anatoly Zlotnik at Los Alamos National Lab for helpful discussions and useful feedback.

\bibliographystyle{elsarticle-num} 
\bibliography{Gas_Dynamics}

\end{document}